\documentclass[12pt,preprint]{aastex}

\newcommand{\gcs}{globular clusters}
\newcommand{\gc}{globular cluster}

\slugcomment{Submitted to AJ}

\shorttitle{Orbital Period of M5 V101}
\shortauthors{Neill et al.}

\begin{document}


\title{The First Orbital Period for a Dwarf Nova in a 
	Globular Cluster: V101 in M5}


\author{James D. Neill\altaffilmark{1}}
\affil{Astronomy Department, Columbia University, New York, NY 10025}
\email{neill@astro.columbia.edu}

\author{Michael M. Shara}
\affil{American Museum of Natural History, 79th and Central Park West
	New York, NY, 10025}
\email{mshara@amnh.org}

\author{Adeline Caulet}
\affil{Calypso Observatory Kitt Peak, 950 North Cherry Avenue, 
	Tucson, AZ, 85726}
\email{caulet@calypso.org}

\and

\author{David A. H. Buckley}
\affil{South African Astronomical Observatory, PO Box 9 Observatory 7935, 
	Cape Town, South Africa}
\email{dibnob@saao.ac.za}


\altaffiltext{1}{Guest user of the Canadian Astronomy Data Centre, which is
operated by the Dominion Astrophysical Observatory for the Canadian National
Research Council's Herzberg Institute of Astrophysics}


\begin{abstract}

We report the first orbital period determination for a Dwarf Nova (DN)
in a \gc: V101 in M5 has a period of 5.796 $\pm$ 0.036 hours.  We derived
this period from I-band photometry acquired with the Calypso Observatory
High Resolution Camera operating with tip-tilt adaptive optics correction.
Observations from the South African Astronomical Observatory in the V-band
were also analyzed and exhibit a periodic signal of the same period.
This orbital period suggests that V101 has a secondary of mid to late K
spectral type with M$_V = +8.2$ $\pm$ 0.5.  The predicted spectral type
is consistent with previous spectral observations in quiescence which
show a fairly red continuum.  From the observed minimum brightness of V
= 22.5, we derive a distance modulus of $(m - M)_V$ = 14.3 $\pm$ 0.5 to
the DN which supports V101's membership in the \gc\ M5.  Measurement of the
ellipsoidality effect indicates that the orbital plane of the V101 system
is moderately inclined, but not enough to exhibit eclipses.

\end{abstract}


\keywords{binaries: eclipsing --- novae, cataclysmic variables --- globular
clusters: individual (M5)}


\section{Introduction}

Close binaries in \gcs\ contain a significant fraction of the total
binding energy of the clusters.  This makes them critically important
in unraveling the dynamical evolution of their host clusters.  In order
to place constraints on cluster evolution theory, one must determine the
binding energy of the binaries in a cluster, i.e. their orbital periods
and component masses.  Determining the evolutionary states of cluster
binaries is also important, and this also demands component masses and
periods.  The high stellar densities in the cores of many \gcs, where
most of the close binaries reside, make this an observational challenge.
UV and X-ray observations have improved our population census of cluster
close binaries (Knigge et al. 2002; Grindlay et al. 2001; Pooley et al.
2002), but without
knowing the periods of these new binaries we cannot calculate the binding
energy they contain or constrain their evolutionary state.

The cataclysmic variable (CV) V101 in the \gc\ M5 has many properties that
make it an ideal candidate for period determination.  It lies 280" from
the center of M5 (10 core radii) and is relatively uncrowded compared
to the objects near the core of a \gc\ (see Figure~\ref{cimi_fig}).  It is 
classified as a dwarf nova (DN), a particularly
well studied class of CV for which empirical relations exist relating
the period to the luminosity, spectral type, and mass of the secondary.

The literature on M5 V101 is not extensive.  It was first reported as a
possible SS Cyg type variable star by \citet{oos41}.  He reported seven
observations, including two pairs of measurements in outburst separated
by 66 days in 1934.  Since then \citet{mar81} performed the initial
identification of V101 in quiescence and obtained a low resolution spectrum 
showing strong, broad Balmer and He I emission lines, confirming the 
identification and the DN classification.  \citet{sha87}
reported B-band observations in outburst and quiescence and predicted a
fairly long orbital period (P $\sim$ 11 hr) based on the minimum observed
magnitude (V = 22.5, Kukarkin \& Mironov 1970) and the duration and rise
and fall time of the outburst.  Spectral observations of V101 in outburst
were published by \citet{nay89} and showed H-$\alpha$ in absorption
with a fairly red continuum.  They also showed a radial velocity curve
covering 1.75 hrs from which they concluded that the orbital period must
be longer than twice their sampling interval (P $\geq$ 3.5 hr, see their
Figure 2).  \citet{sha90} published the highest resolution spectrum to
date of V101 in quiescence and noted that the high velocity width of
the Balmer lines hinted that the system might be highly inclined and
eclipses might be observable.

The possibility of using these eclipses to determine the orbital period
for M5 V101 motivated this study.  The availability of many nights on
the Calypso Observatory 1.2m telescope using the high resolution camera
allowed us to achieve the photometric accuracy and time sampling required
to determine the orbital period.  We now describe the photometric analysis
and the method for finding the period and present the physical properties
of V101 derived from our observations.

\section{Observations}

\subsection{The Calypso Telescope Observations}

Since these are the first scientific results from the Calypso Telescope,
we present a brief description of the telescope and camera.  A more
detailed description will be published elsewhere.

The Calypso Telescope is a 1.2m telescope of Ritchey-Chretien design
on a computer controlled altitude-azimuth mounting located on Kitt
Peak, Arizona.  Two instruments are mounted at the Naysmith foci: a
high resolution camera (HRCAM) with tip/tilt adaptive correction and a
field of view of 85", and a non-adaptive camera (WFCAM) with a 10' field
of view.  The aperture of the telescope was matched to the atmospheric
cell size during the best quartile of seeing at the site allowing the
tip/tilt to theoretically correct 87\% of the atmospheric distortion
on the optical axis.  The optical components are figured so that the
total wavefront error at the focal plane, including all error budgets,
is less than $1/17$th of a wavelength at 5500\AA.  The telescope is 
mounted on a 10.2 meter high pier with
an enclosure that rolls completely away, allowing it to operate in the
open air.  The site is placed so the prevailing winds during the best
seeing reach the telescope unhindered by any ground obstruction providing
a laminar flow of air over the telescope.

\begin{figure}
\figurenum{1a}
\plotone{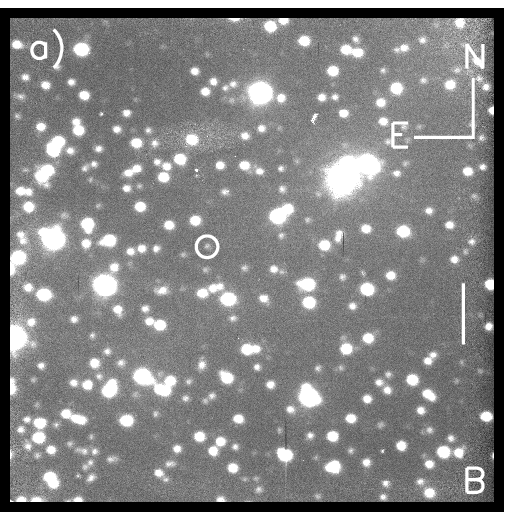}
\caption{Calypso B-band image of V101 produced by coadding three 600 second
HRCAM exposures taken on JD 2451994 (26 March 2001).  The scale bar on the 
right is 10 arcseconds long, and V101 is circled.  The point spread function
has a full width at half maximum of 0.8 arcseconds.\label{cimb_fig}}
\end{figure}

\begin{figure}
\figurenum{1b}
\plotone{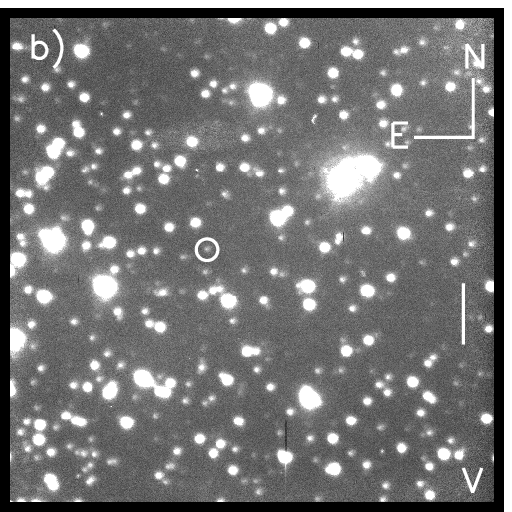}
\caption{Calypso V-band image of V101 produced by coadding three 600 second
HRCAM exposures taken on JD 2451993 (25 March 2001).  The scale bar on the 
right is 10 arcseconds long, and V101 is circled.  The point spread function
has a full width at half maximum of 0.8 arcseconds.\label{cimv_fig}}
\end{figure}

\begin{figure}
\figurenum{1c}
\plotone{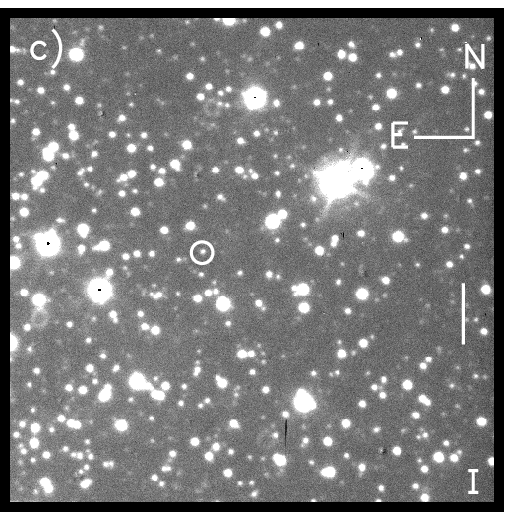}
\caption{Calypso I-band image of V101 produced by coadding three 600 second
HRCAM exposures taken on JD 2452051 (22 May 2001).  The scale bar on the 
right is 10 arcseconds long, and V101 is circled.  The point spread function
has a full width at half maximum of 0.6 arcseconds.\label{cimi_fig}}
\end{figure}

The HRCAM was used for this study of V101.  It uses a single Loral
2048 square CCD with 15$\mu$ pixels.  Its native resolution is 0.04"
per pixel in order to over-sample the point spread function (PSF) at the
best seeing measured at the site of 0.25".  However, the average seeing
ranges between 0.6 and 0.8", as measured in the I-band.  Therefore, the
CCD was binned 4x4 (giving a pixel size of 0.16") to better accommodate
these atmospheric conditions during the observing campaign.  Since the
secondary component of V101 was thought to be a late-type main-sequence
(M-S) star, we chose the I-band to maximize the chances that ellipsoidal
variations would be visible.  We used 10 minute exposures back-to-back to
allow us to sample the light curve densely enough to detect low amplitude
eclipses or ellipsoidal variations.  This exposure time gave us a limiting
magnitude of I $\sim$ 21.5 under typical conditions.  Table \ref{obs_tab}
presents the log of I-band observations for the Calypso data.  Figures
\ref{cimb_fig} - \ref{cimi_fig} present three finder charts for V101
in the B, V, and I-bands taken with the HRCAM on Calypso.

\subsection{South African Astronomical Observatory Observations}

Observations taken in 1995 with the TEK4 CCD on the South African
Astronomical Observatory (SAAO) 1.9m telescope in the V-band were
also analyzed for this study.  For five nights, V101 was observed
with 15 minute exposures back-to-back over the entire night yielding
a limiting magnitude of V $\sim$ 22.7 under typical conditions.
Table \ref{obs_tab} presents the log of observations for the SAAO
V-band data.  These observations proved valuable for confirming the
orbital period seen in the Calypso data (see below).

\section{Reduction and Photometry}

All frames were reduced in the standard way to remove instrumental
artifacts.  In order to minimize the impact of cosmic rays, each image
was shifted to a standard reference position and coadded in the following
way: first, all images were ordered in time sequence, then a running set
of three images was coadded from the beginning to the end of the night
incremented by one image at a time.  This produced a smoothing of the
light curve but kept the time sampling interval at roughly the exposure
time of the individual images (plus read time).

The coadded frames were photometered using the APPHOT package in
IRAF \citep{tod86}.  A set of isolated, well-exposed stars near
V101 with low variability ($\leq$ 0.02 mag) was used to tie all
the epochs together onto the same instrumental magnitude system.
Calibration was achieved by comparing our instrumental magnitudes
with Peter Stetson's photometry of M5\footnote{available at
http://cadcwww.dao.nrc.ca/cadcbin/wdb/astrocat/stetson} \citep{ste00}
in the I and V-bands.  Due to the small fields of the Calypso and SAAO
detectors the calibration was boot strapped to the Stetson standards
through intermediate wider-field images taken under photometric
conditions.  A wide field I-band image of M5 taken with the 8K mosaic
camera on the Hiltner 2.4m telescope at the MDM observatory on 01 June
2001 by JDN was used to bootstrap the Calypso photometry.  The SAAO
photometry was bootstrapped using a V-band image kindly taken for us
by Ron Downes with the T1KA camera on the 2.1m telescope at Kitt Peak
National Observatory on 28 May 1998.  In each case and at each step at
least 25 stars were used and a final absolute photometric calibration of
better than 0.1 mag was achieved for all photometry.  The final calibrated
I-band magnitudes are presented in Table~\ref{cal_phot_tab}, and the final
calibrated V-band magnitudes are presented in Table~\ref{sa_phot_tab}.

\begin{figure}
\figurenum{2}
\plotone{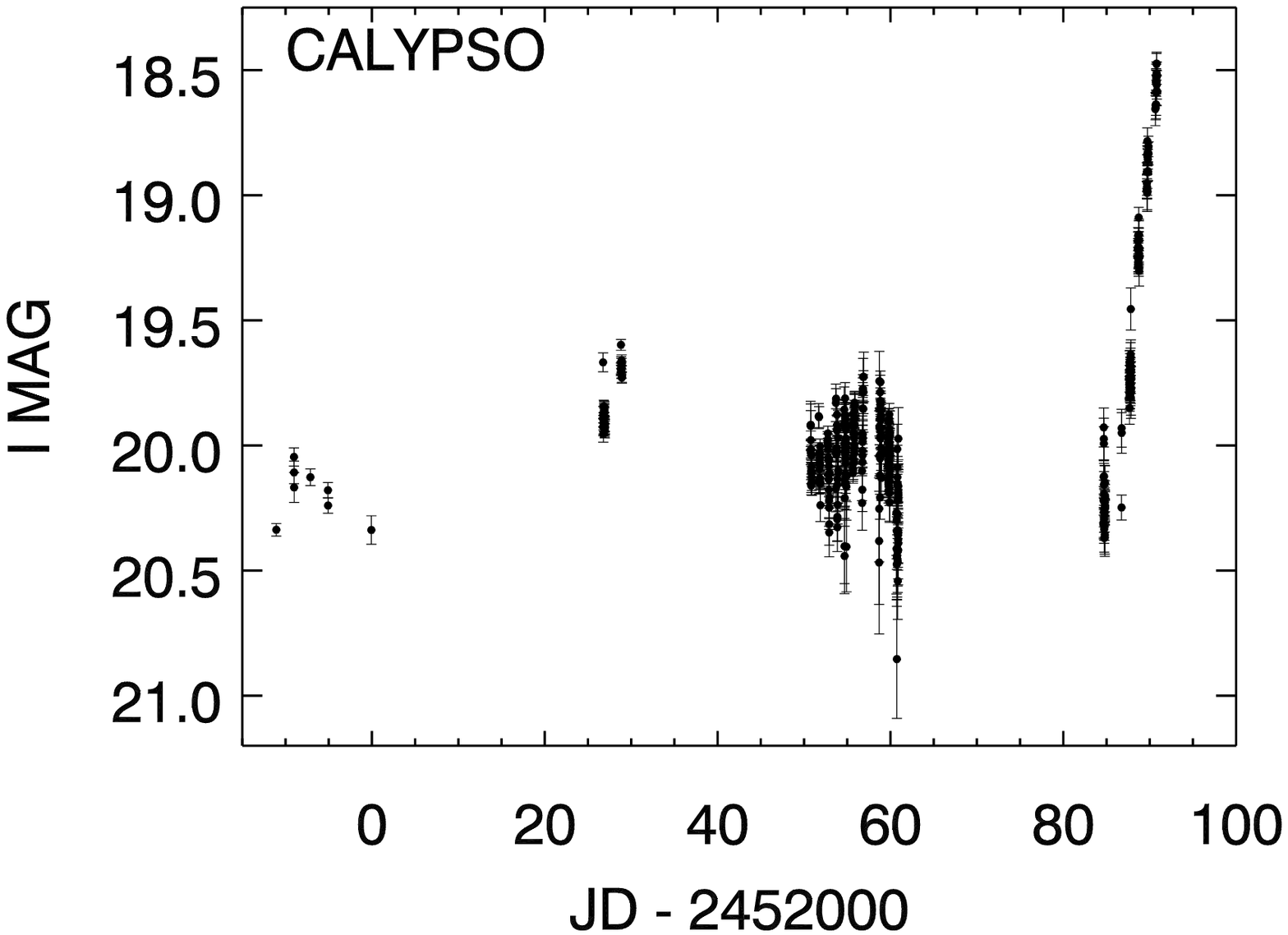}
\caption{Calypso photometry summarized showing the individual I-band
observations.
\label{cphot_sum_fig}}
\end{figure}

\begin{figure}
\figurenum{3}
\plotone{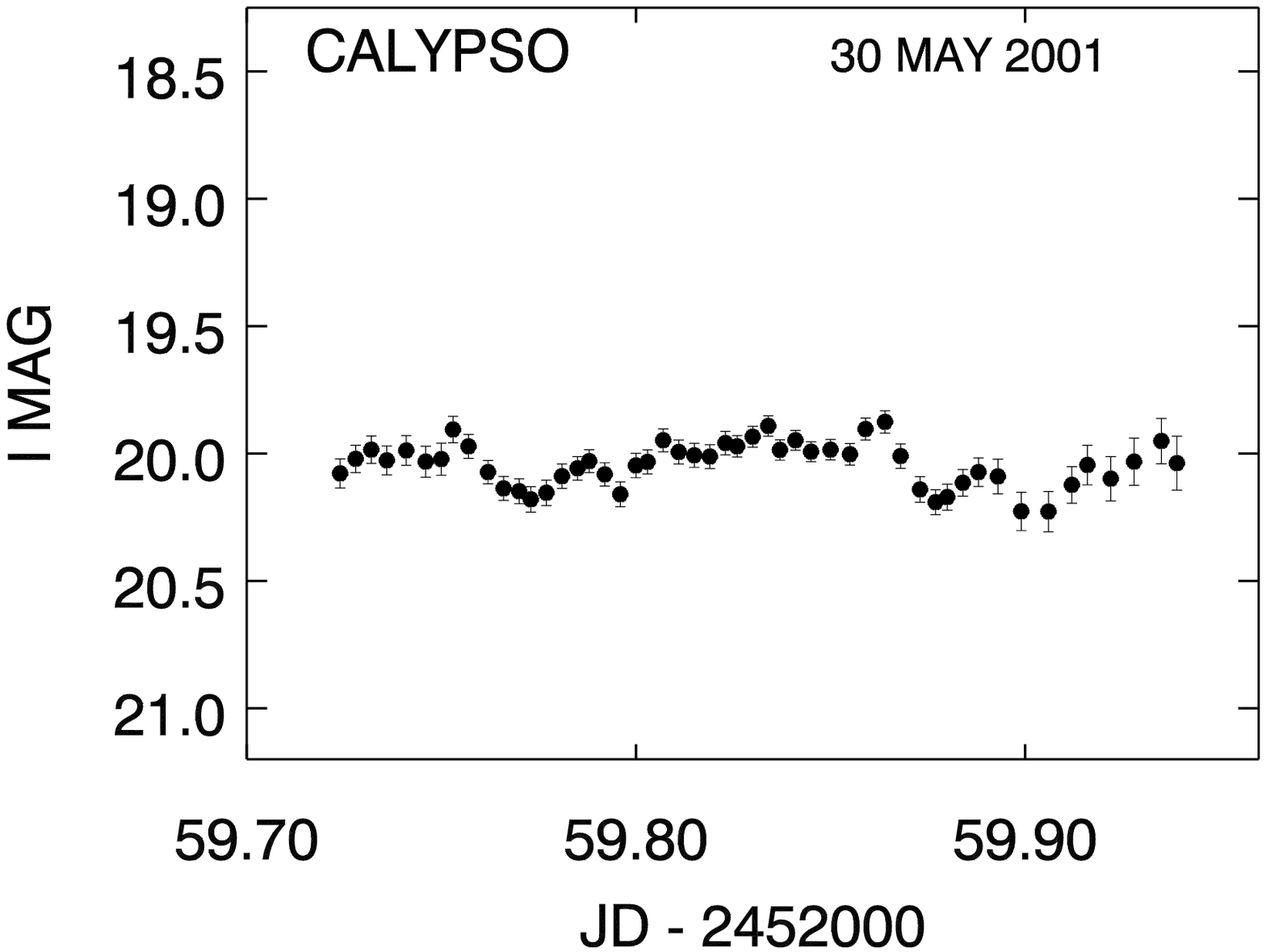}
\caption{One night from the Calypso observations during quiescence showing
the `flickering' typical of CVs.
\label{cphot_one_fig}}
\end{figure}

\begin{figure}
\figurenum{4}
\plotone{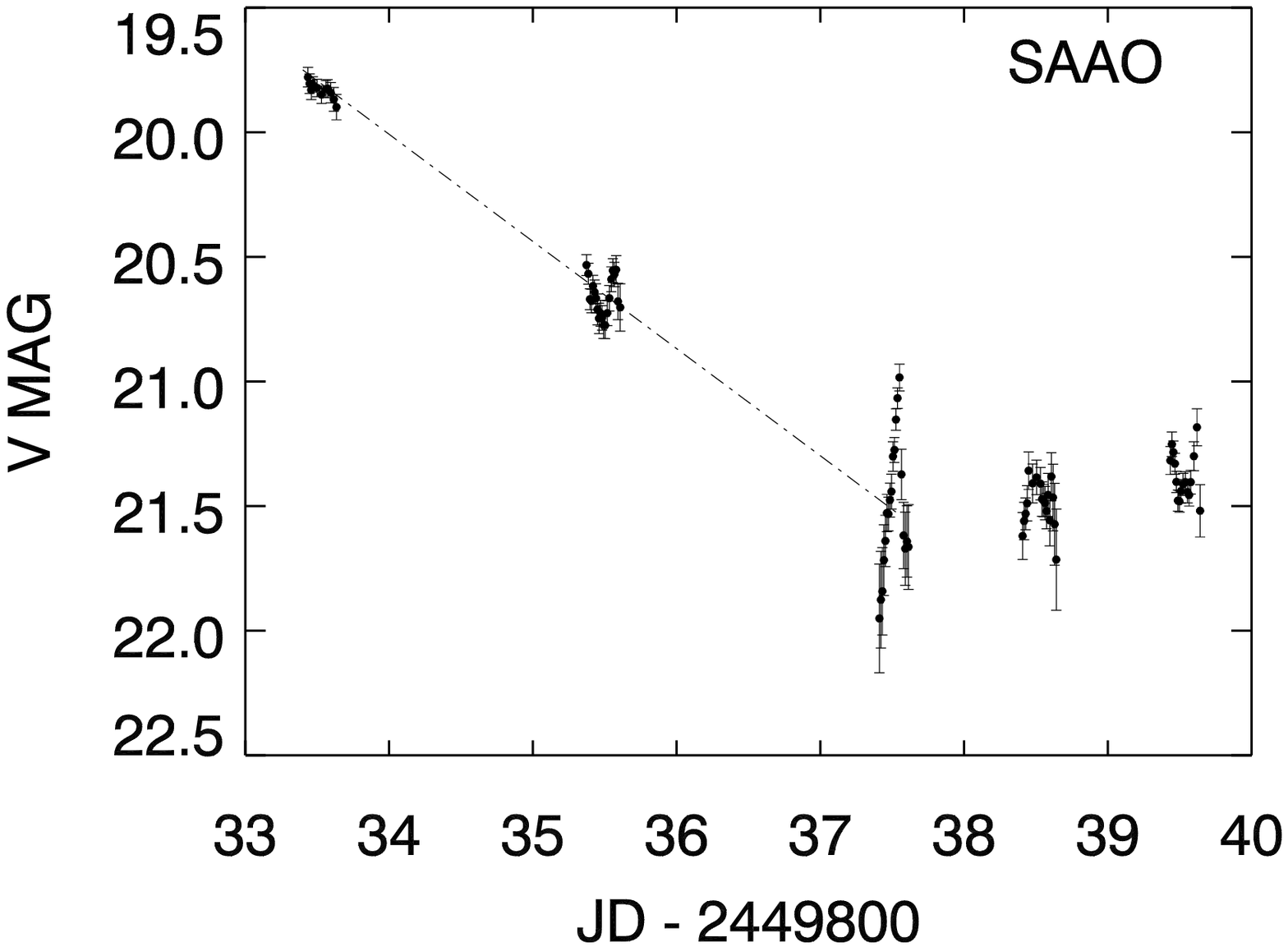}
\caption{SAAO photometry summarized showing the individual V-band
observations.  The empirical relation between orbital period and 
outburst decline rate from equation 3.5 of \citet{war95} overplotted as a 
dot-dashed line (see \S~\ref{sec_orb}).
\label{sphot_sum_fig}}
\end{figure}

\begin{figure}
\figurenum{5}
\plotone{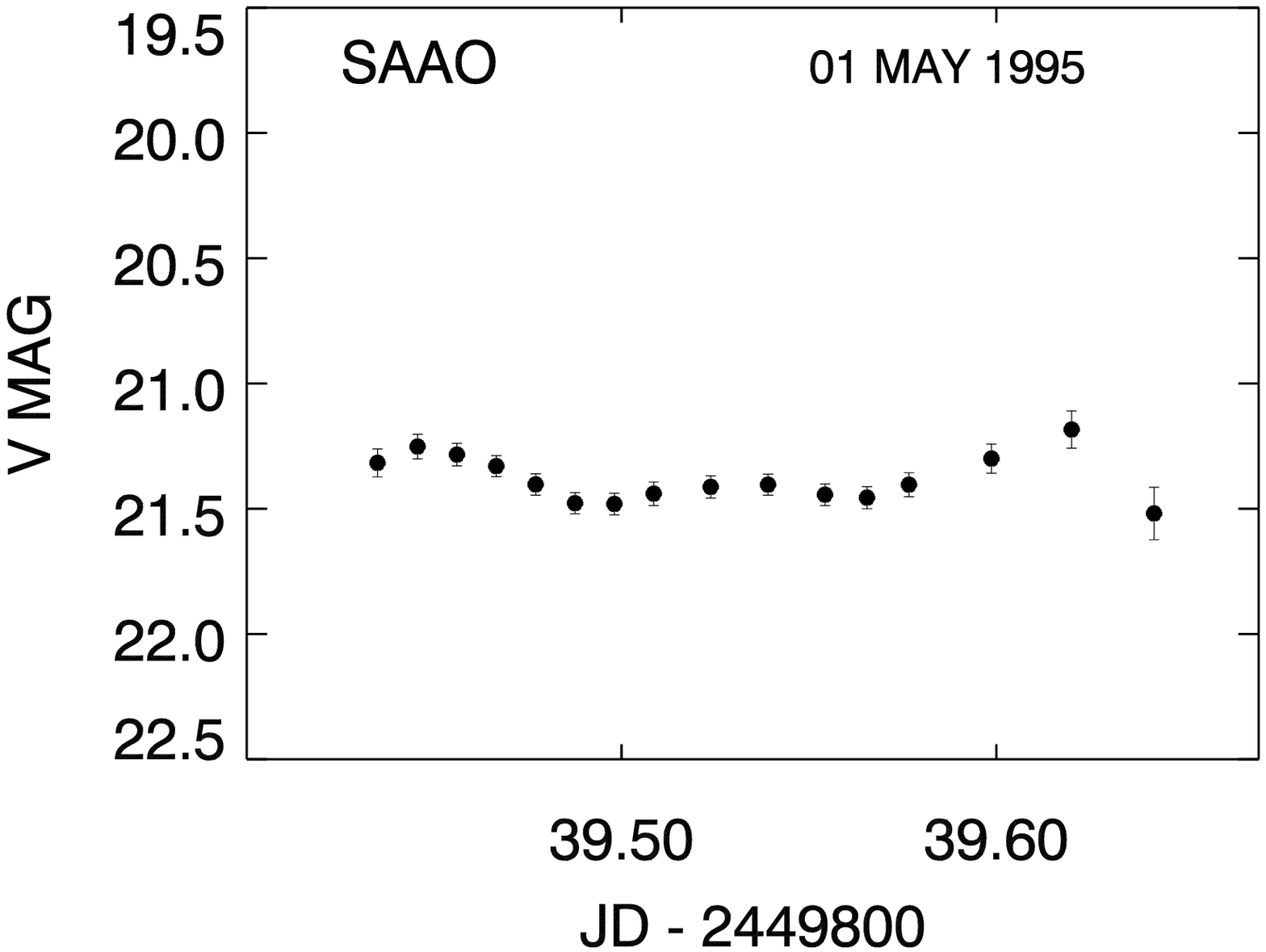}
\caption{One night from the SAAO observations during quiescence.
\label{sphot_one_fig}}
\end{figure}

The I-band photometry is summarized in Figure~\ref{cphot_sum_fig} with
a single night during quiescence shown in Figure~\ref{cphot_one_fig} to
illustrate a typical night's I-band light curve.  The V-band photometry
is summarized in Figure~\ref{sphot_sum_fig} and a single night's V-band
observations during quiescence are shown in Figure~\ref{sphot_one_fig}.
Figure \ref{cphot_sum_fig} shows what may be the beginning of an outburst
on JD 2452028 and one well observed outburst rise starting at JD 2452087.
Figure \ref{sphot_sum_fig} shows a decline from outburst to a V magnitude
of 22.0 on JD 2449837.

\section{Orbital Period Analysis}

To derive the orbital period we used the algorithm of \citet{sca82} and
\citet{hor86}, accelerated by the technique described in \citet{pre89}.
The calibrated magnitudes were converted to flux units for the period
analysis.  Residual long term trends were removed by taking each night,
calculating the mean flux for the night, and subtracting this mean flux
from the individual fluxes from the night.

\subsection{Calypso Data \label{sec_calypso}}

\begin{figure}
\figurenum{6}
\plotone{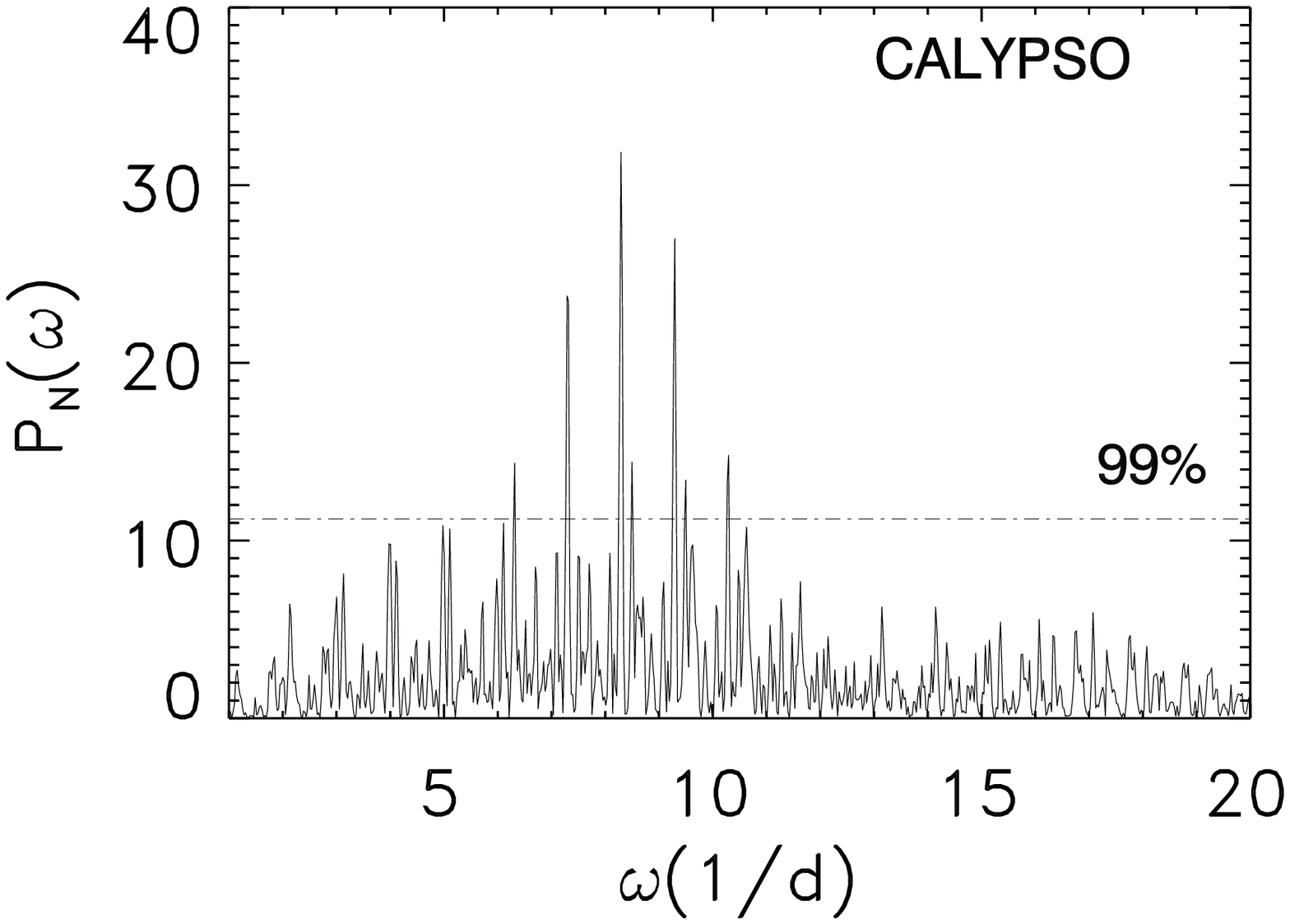}
\caption{Periodogram generated from the Calypso data with errors $\leq$
0.2 magnitudes.  The line of 99\% confidence (false alarm probability
= 10$^{-2}$) is indicated as a dot-dashed line.  The largest peak
is at $\omega$(1/d) = 8.281 $\pm$ 0.026 (P = 2.898 $\pm$ 0.009 h)
and has a false alarm probability of 10$^{-10.9}$.  All other other
significant peaks have corresponding peaks in the alias periodogram
shown in Figure~\ref{cfake_fig}
and are aliases due to the time sampling of the Calypso data (see
\S~\ref{sec_calypso}).\label{cper_fig}}
\end{figure}

We used all data with errors $\leq$ 0.2 magnitudes to generate the
periodogram shown in Figure~\ref{cper_fig}.  The most significant peak
is at $\omega$(1/d) = 8.281 $\pm$ 0.026 (P = 2.898 $\pm$ 0.009 h).
The half-width of the periodogram peak at 85\% of its peak value was 
used to derive all frequency error estimates in this paper.

If the periodic signal is due to the ellipsoidality effect \citep{boc79}
then the period found in the periodogram is half the orbital period
because there are two modulations per orbit.  In the I-band this would
be especially true since the low mass M-S secondary would be prominent.
An odd-even effect, where one modulation is deeper than the other, is
expected because the two sides of the secondary are unequally luminous
either due to gravity darkening or heating by the primary.  This, plus
the radial velocity curve published by \citet{nay89} which indicates
an orbital period $\geq$ 3.5~h, motivated us to explore both the peak
frequency from our periodogram and half this frequency: $\omega$(1/d)
= 4.140 $\pm$ 0.026 (P = 5.796 $\pm$ 0.036 h).

\begin{figure}
\figurenum{7}
\plotone{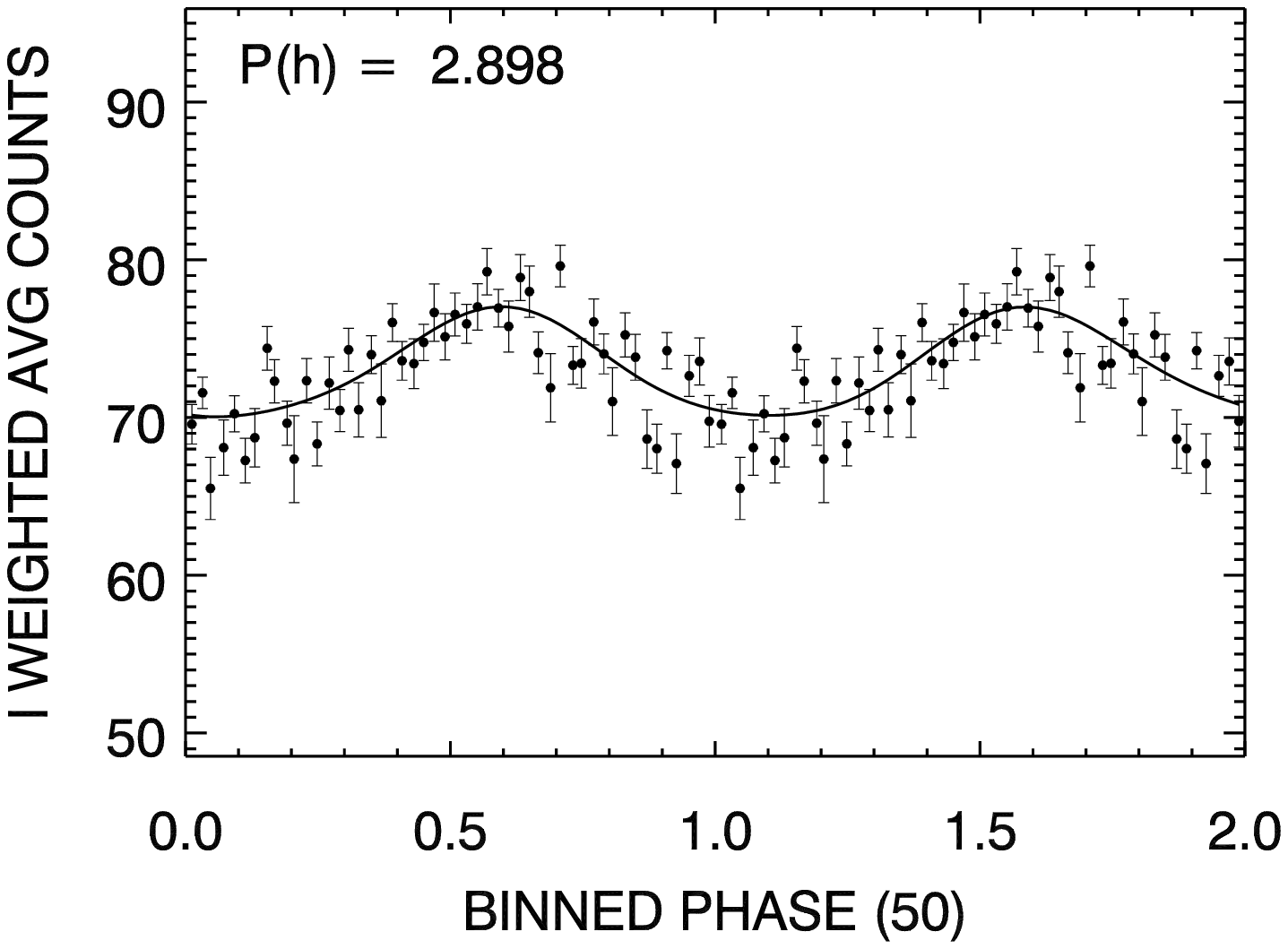}
\caption{ Phase diagram generated from error weighted counts in 50 phase
bins using the period of the largest peak in the Calypso periodogram.
The solid line is the fit of equation~(\ref{eq0}) to the phase points and 
is used 
to generate the alias periodogram shown in Figure~\ref{cfake_fig}. The fit
has a reduced $\chi^2$ of 2.15.
\label{chpha_fig}}
\end{figure}

\begin{figure}
\figurenum{8}
\plotone{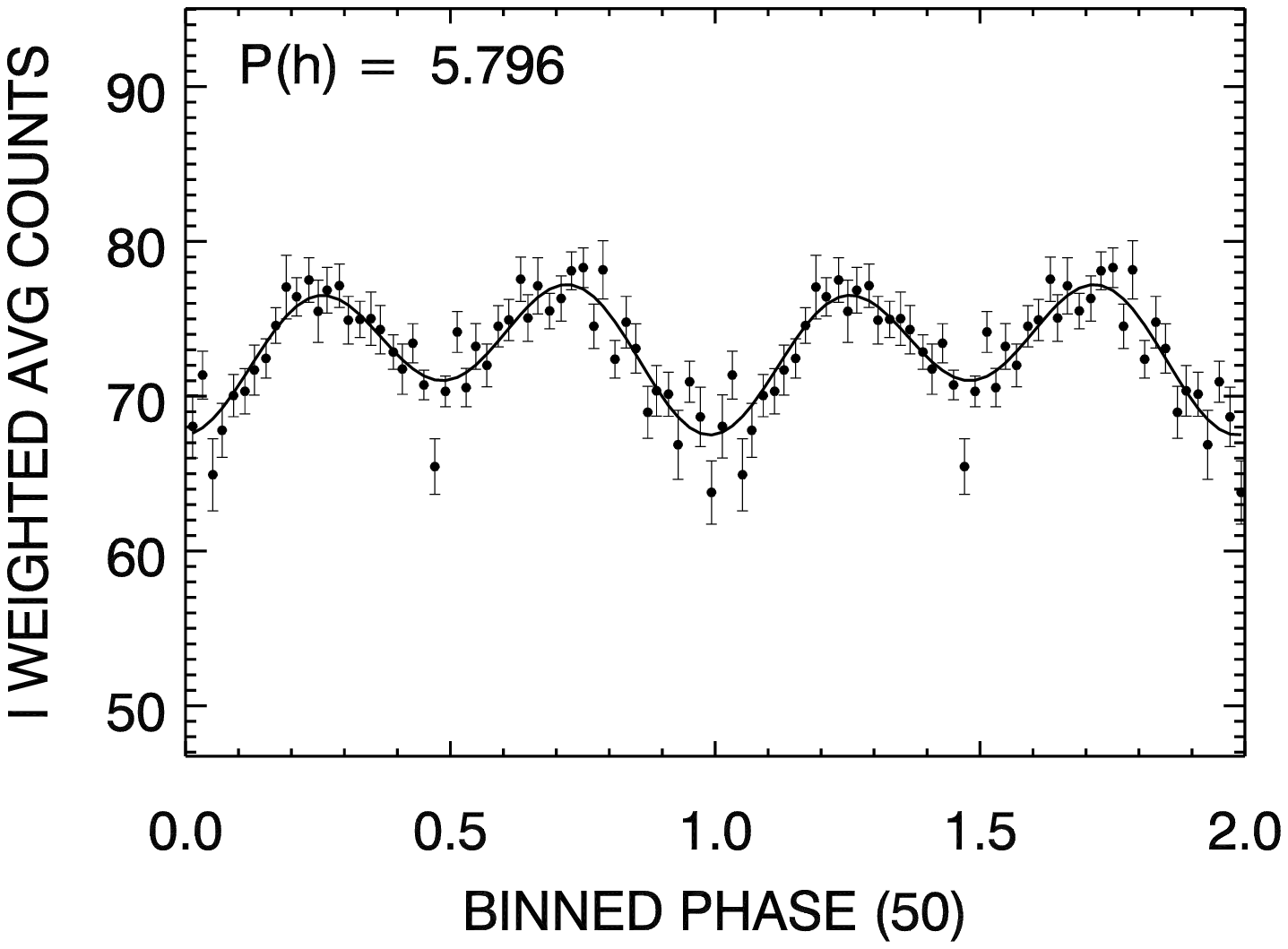}
\caption{ Phase diagram generated from error weighted counts in 50 phase
bins using twice the period of the largest peak in the Calypso periodogram.
The solid line is the fit of equation~(\ref{eq0}) to the phase points and is used 
to analyze the orbital inclination of V101 (see \S~\ref{sec_inc}).  The fit
has a reduced $\chi^2$ of 1.25.
\label{cfpha_fig}}
\end{figure}

We produced the binned phase diagrams shown in Figures~\ref{chpha_fig}
and \ref{cfpha_fig} using the error weighted mean counts within each
of 50 phase bins.  Figure~\ref{cfpha_fig} clearly shows the expected
odd-even effect of two unequal modulations per orbital period.  We fit
both sets of phase data with a periodic function of the form:
\begin{equation}
	y = a + b\cos(c\phi + d) + e\sin(f\phi + g),
\label{eq0}
\end{equation}
where $\phi$ is the phase, and plotted them on the phase data.
The reduced $\chi^2$ of the fit in Figure~\ref{chpha_fig} is 2.15,
while that in Figure~\ref{cfpha_fig} is 1.25, which supports the longer
orbital period.

\begin{figure}
\figurenum{9}
\plotone{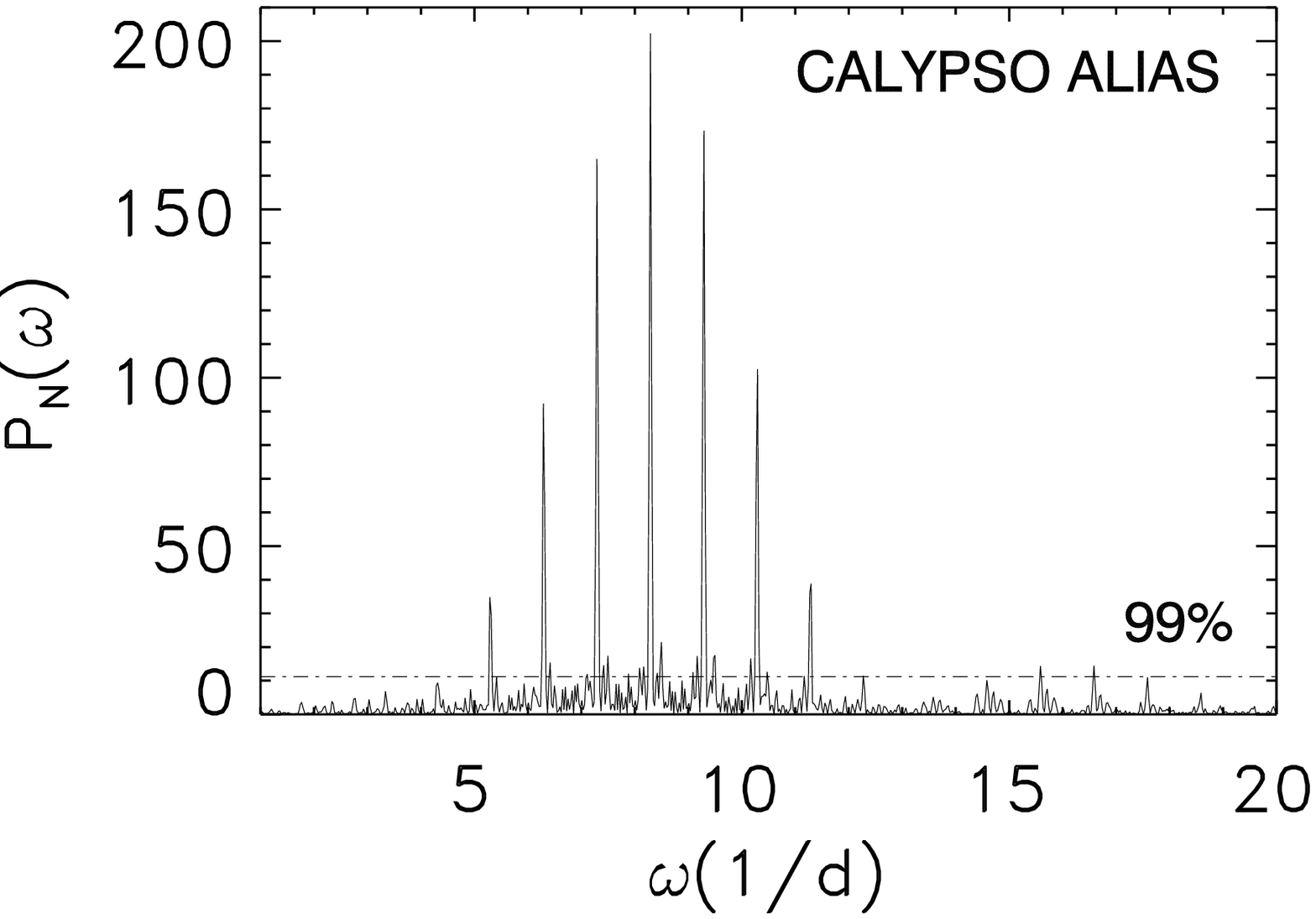}
\caption{Periodogram generated from the fit in Figure~\ref{chpha_fig}
sampled the same way as the Calypso observations and
showing the alias peaks.\label{cfake_fig}}
\end{figure}

The fit of equation~(\ref{eq0}) to the phase data was sampled the same way
as the Calypso data and used to generate the alias periodogram shown in
Figure~\ref{cfake_fig}.  All significant peaks from Figure~\ref{cper_fig}
have a corresponding peak in Figure~\ref{cfake_fig} and, apart from the
highest one, are aliases due to our time sampling.

\subsection{SAAO Data}

\begin{figure}
\figurenum{10}
\plotone{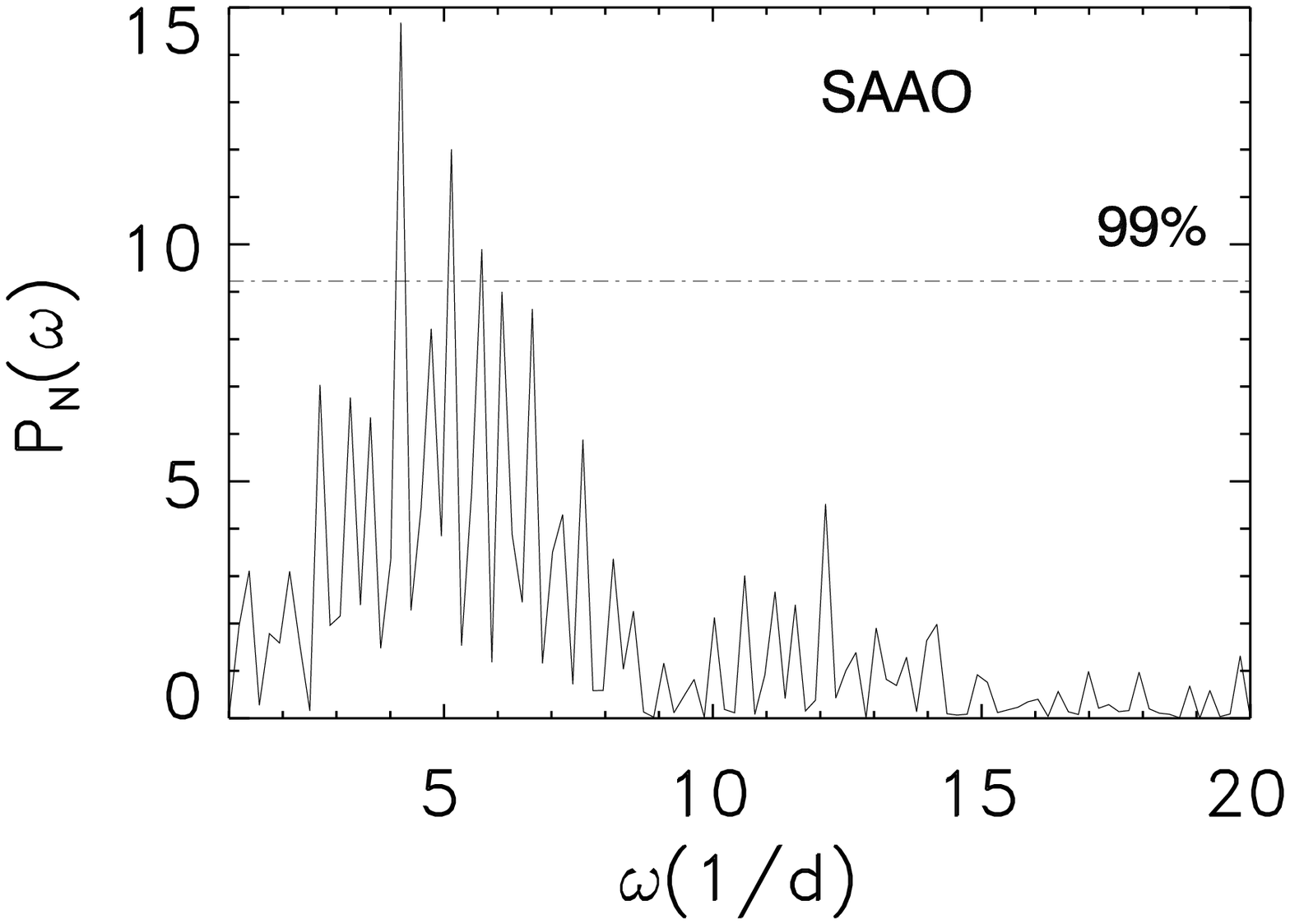}
\caption{Periodogram generated from the SAAO data.  The largest peak is 
at $\omega$(1/d) = 4.20 $\pm$ 0.19 (P = 5.72 $\pm$ 0.25 h) and has a 
false alarm probability of 10$^{-4.5}$.
\label{sper_fig}}
\end{figure}

We applied the same techniques to the V-band data taken at the SAAO.
We assumed that the orbital modulations would be present during the
decline from outburst and used the run of observations from JD 2449833
to JD 2449839.  Figure \ref{sper_fig} presents the periodogram derived
from these data.

\begin{figure}
\figurenum{11}
\plotone{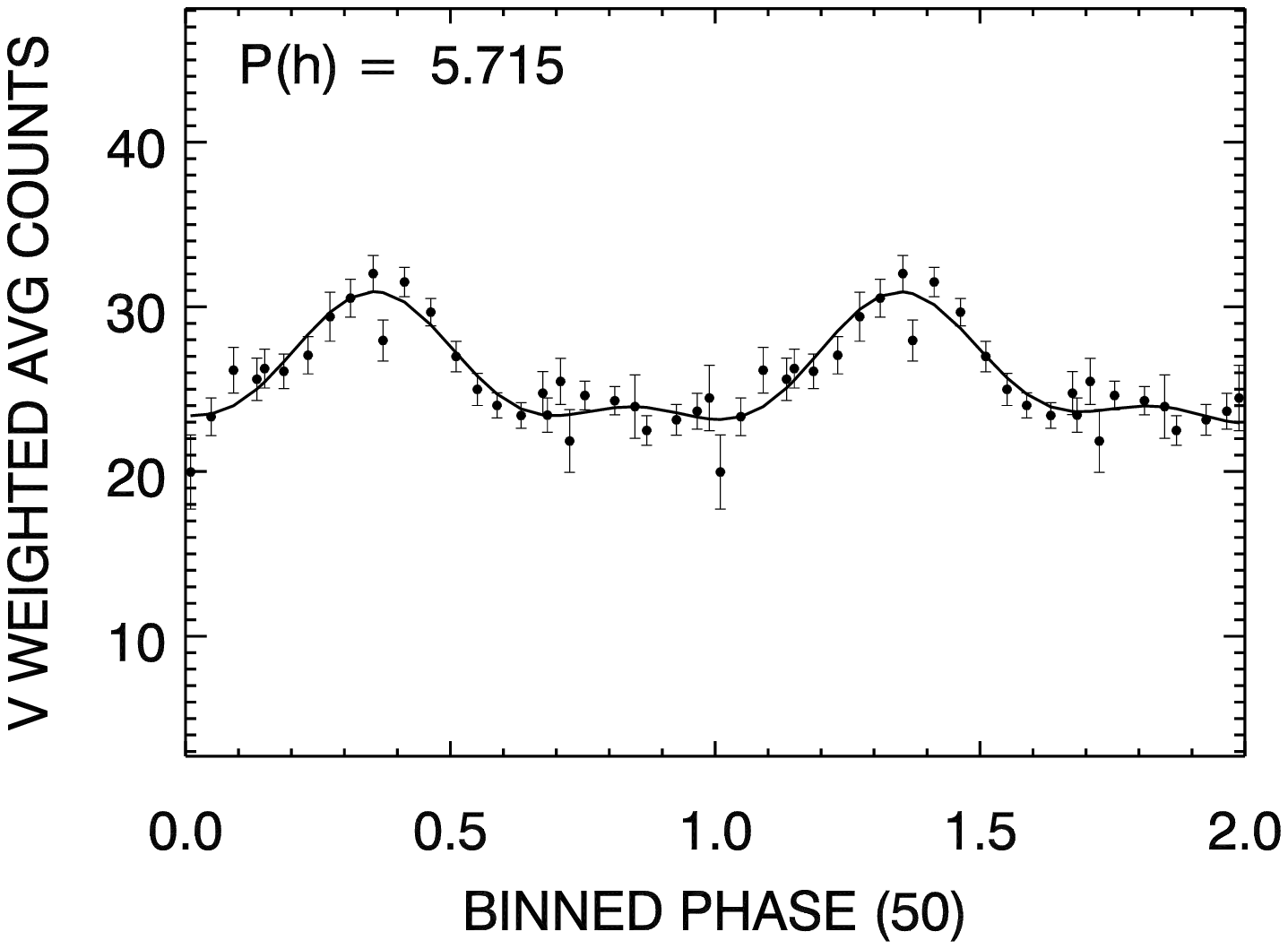}
\caption{Phase diagram generated from error weighted counts in 50 phase
bins using the period of the largest peak in the SAAO periodogram.
The solid line is the fit of equation~(\ref{eq0}) to the phase points and is used 
to generate the alias periodogram shown in Figure~\ref{sfake_fig}. The fit
has a reduced $\chi^2$ of 1.09.
\label{sfpha_fig}}
\end{figure}

While this periodogram is not as clean as the periodogram derived from
the Calypso data, its most significant peak at $\omega$(1/d) = 4.20 $\pm$
0.18 (P = 5.72 $\pm$ 0.25 h) does support the 5.796 hour period from the
Calypso data.  The binned phase diagram using the weighted mean counts
within each of 50 phase bins is presented in Figure \ref{sfpha_fig}.

\begin{figure}
\figurenum{12}
\plotone{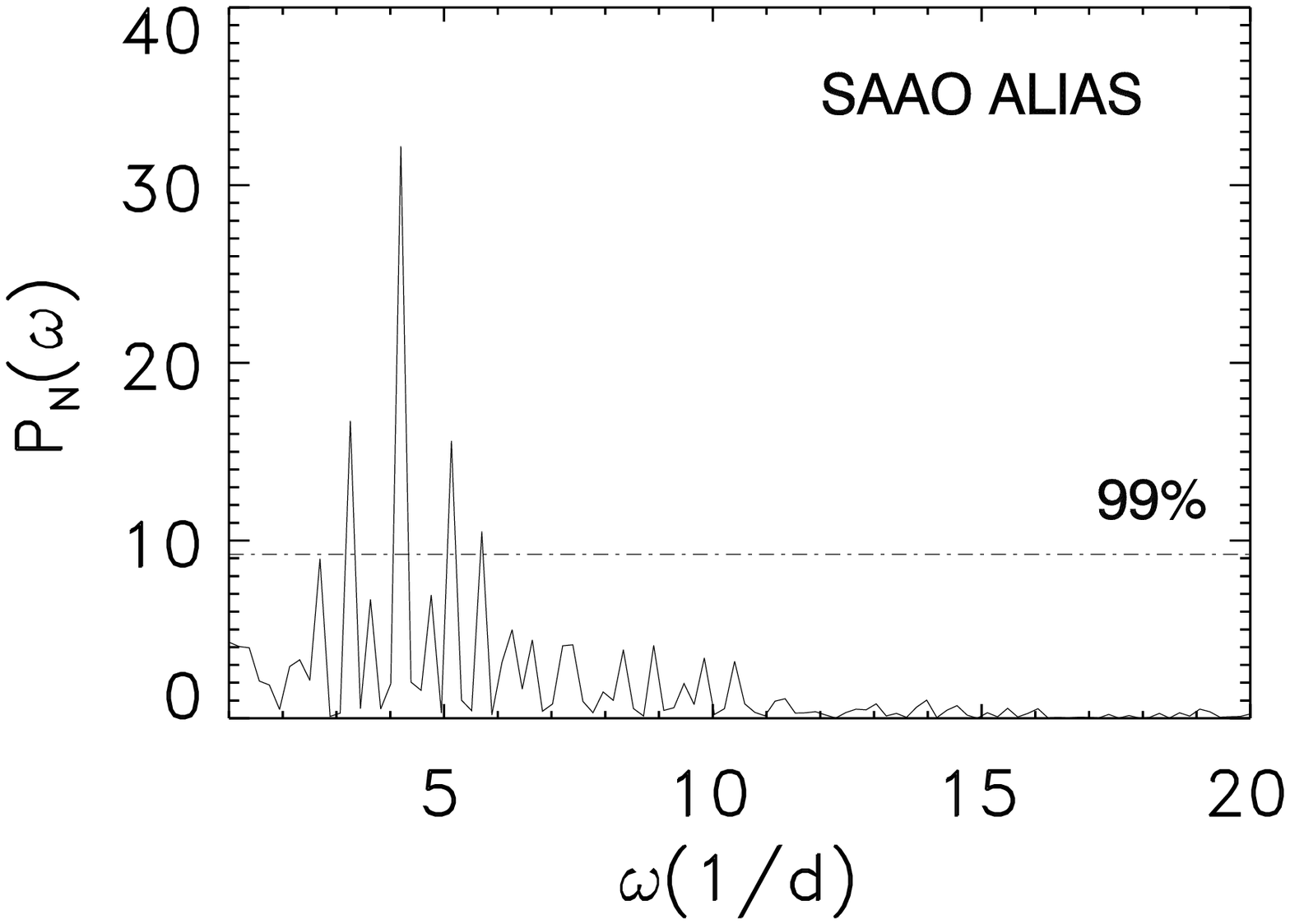}
\caption{Periodogram generated from the fit in Figure~\ref{sfpha_fig}
sampled the same way as the SAAO observations and 
showing the alias peaks.
\label{sfake_fig}}
\end{figure}

We also fit this phase diagram with equation~(\ref{eq0}) in order to
investigate aliases in the periodogram.  The fit is presented as the
solid line in Figure~\ref{sfpha_fig}.  The alias test periodogram is
presented in Figure \ref{sfake_fig}.  Most of the significant peaks can
be explained as aliases from our time sampling.

\section{Interpretation}

\subsection{The Orbital Period of V101\label{sec_orb}}

The phase diagram presented in Figure \ref{cfpha_fig} shows the odd-even
phenomenon from the ellipsoidality effect, caused by the changing
geometry of the distorted secondary during the orbital cycle and the
unequally luminous halves of the secondary.  This effect has a small
amplitude and only with many observations can the superposed stochastic
variations from the accretion process be averaged out.

Figure 2.45 from \citet{war95}, showing the observed relationship
between spectral type and P$_{orb}$, places the secondary of V101 in the
range of spectral types from K5 to M0.  This is consistent with spectral
observations published by \citet{mar81}, \citet{nay89}, and \citet{sha90}
which show a red continuum in quiescence.

The phase diagram for the SAAO V-band data presented in Figure
\ref{sfpha_fig} shows a modulation of the same period.  We expect the
secondary to be much fainter in the V-band, since the secondary is of
spectral type K5 to M0 with $V - I \simeq 2.2$, and so we see only one
modulation per orbit, perhaps due to the changing visibility of the
primary star or the changing aspect of the accretion disk.

As a further consistency check, we compare the decay from outburst shown
in Figure \ref{sphot_sum_fig} with equation 3.5 from \citet{war95},
which relates the outburst decay timescale, $\tau_d$, to P$_{orb}$:
\begin{equation}
	\tau_d = 0.53\ P_{orb}^{0.84}(h)\ {\mathrm d}{\mathrm a}{\mathrm y}\ {\mathrm m}{\mathrm a}{\mathrm g}^{-1}.
\label{eq2}
\end{equation}
We have overplotted this relation for an orbital period of 5.796 hours
on Figure \ref{sphot_sum_fig} as a
dot-dashed line which follows the decline well.  This orbital period 
places V101 well above the period gap for CVs and at the high end
of the distribution for CVs above this gap \citep{war95}.

\subsection{The Orbital Inclination of V101 \label{sec_inc}}

Having a constraint on the spectral type of the secondary allows us to
use the light curve to explore the orbital inclination of the system
using the tables in \citet{boc79}, assuming that the modulations shown
in Figure~\ref{cfpha_fig} are purely due to the ellipsoidality effect.
The fit shown in Figure~\ref{cfpha_fig} as the solid line and the average
deviation from this fit was used to determine the amplitude of the
variation ($A = 0.114 \pm 0.020$~mag), the difference between the two
minima ($\Delta m = 0.055 \pm 0.021$~mag), and their corresponding errors.

Knowing that the secondary is a K5 to M0 dwarf constrains the effective
temperature which, combined with the filter bandpass and the fact that the
envelope of the secondary star is convective, allows us to use Figure 2
from \citet{boc79} to determine the gravity darkening coefficient, $\beta
\simeq$ 0.4.  The limb darkening coefficient is also determined by the
effective temperature and using Figure 17.6 from \citet{gra76} we get u
$\simeq$ 0.5.  We know that the secondary is filling its Roche lobe so
we can set the Roche lobe filling factor to be $\mu$ = 1.  We can also
assume that the mass ratio, q $= M_p/M_s$ is in the range 1.6 - 2.3, by
using the mass of a K7 dwarf for the secondary star mass, $M_s \simeq$
0.6 M$_{\odot}$, and assuming the primary star mass, $M_p$, is in the
range 1.0 to 1.4 M$_{\odot}$ (see \S~\ref{mass_loc}).  Note that the q
used by \citet{boc79} is the inverse of that used traditionally in the
CV literature where q $= M_s/M_p$.

Using these values to examine Table 1 from \citet{boc79} we can constrain
the inclination angle, $i$, by $A$ to be in the range $30^\circ < i
< 60^\circ$.  Using Table 2 from \citet{boc79} we can constrain the
inclination angle by $\Delta m$ to be in the range $50^\circ < i <
90^\circ$.  These two constraints overlap in the range $50^\circ < i <
60^\circ$ and are consistent with the fact that no eclipses are seen.

\subsection{The Distance to V101}

We can now use the empirical relation between orbital period, $P_{orb}$,
and secondary luminosity, $M_V(2)$, to determine the distance modulus
to V101.  Using equation 2.102 from \citet{war95},
\begin{equation}
	M_V(2) = 16.7 - 11.1\ {\mathrm l}{\mathrm o}{\mathrm g}\ P_{orb}(h),
\label{eq1}
\end{equation}
we determine the absolute magnitude of the secondary to be M$_V = +8.2$.
Figure 2.46 in \citet{war95} shows that the scatter in this relation is
about $\pm$ 0.5 magnitudes.  From the minimum published V magnitude of
22.5 \citep{kuk70} and M$_V = +8.2$, we get a distance modulus of $(m -
M)_V =$ 14.3 $\pm$ 0.5, consistent with the distance modulus of M5 of
$(m - M)_V = 14.41 \pm 0.07$ \citep{san96}.

\subsection{The Outburst Period of V101}

If the rise seen in Figure~\ref{cphot_sum_fig} on JD 2452028 is indeed the
beginning of an outburst, then the two outbursts we observe separated
by 60 days with an intervening quiescent period supports the outburst
period first proposed by \citet{oos41} of 66 days.  Many observations of
individual outbursts have been reported (Margon, Downes, \& Gunn 1981;
Shara, Potter, \& Moffat 1987; Naylor et al. 1989).  Using these data to
examine the periodicity of the outbursts shows that no regular period
for the outbursts exists, but a `typical' outburst interval is in the
range of 60 to 66 days.  This is expected behavior for DN outbursts
which don't exhibit strict periodicity.

\subsection{The Mass and Location of V101 \label{mass_loc}}

A recent mass-spectral type study \citep{bar96} concludes that an M0
star has a mass of 0.6 M$_{\odot}$.  The mass-orbital period relation,
equation 2.100 from \citet{war95},
\begin{equation}
  M_1(2) = 0.065\,P_{orb}^{5/4}(h) \;\;\;\;\;\; 1.3 \leq P_{orb}(h) \leq 9,
\label{eq3}
\end{equation}
yields 0.58 M$_{\odot}$ as the secondary mass of a CV with a 5.796 hour
period. These are consistent with the 0.6 M$_{\odot}$ derived earlier
from the system luminosity (and type K5-M0) near minimum.  The white dwarf
mass of V101 must then be $>$ 1 M$_{\odot}$ or so to prevent dynamical
mass transfer, implying a system mass in excess of 1.6 M$_{\odot}$.
This system mass is twice that of the main sequence turnoff in M5,
which should place V101 in the inner one or two core radii of the cluster.

M5 V101 stubbornly refuses to conform to this logic.  It is, in fact,
located 10 core radii from the center of M5. What is it doing out
there? Perhaps the simplest explanation is dynamics.

Stars lead extremely promiscuous lives in clusters, especially near
the centers where stellar densities are highest.  Mate swapping is
commonplace, as are strong, close encounters between binaries and single
stars (e.g. Hurley \& Shara 2002).  These encounters often lead to
the high speed recoils of the emerging binary and single stars. Such
a scenario could place V101 far from the core of M5. It also makes
the intriguing prediction of the existence of a low mass M dwarf  on
the opposite side of M5, considerably farther out than 10 core radii,
escaping the cluster at high speed.

\section{Conclusions}

We conclude that the orbital period for V101 is P = 5.796 $\pm$ 0.036 h.
Using this orbital period to determine a distance modulus yields $(m
- M)_V = 14.3 \pm 0.5$ which supports the membership of V101 in M5.
We conclude that the secondary of V101 is a low mass M-S star whose
spectral type is in the range K5 to M0.  We also conclude that the
orbital inclination is high, but not high enough for the system to
exhibit eclipses.



\acknowledgments

For the generous allocation of observing time for this project as well
as material and intellectual support we are indebted to the Calypso
Observatory Director, Edgar Smith.  We acknowledge the enthusiastic
support of our engineers, Bruce Truax and Frank Scinicariello, who
made this project possible.  JDN would like to acknowledge many useful
conversations with Joe Patterson concerning methods for analyzing the
periodic signal in photometric data.  MMS and DAHB gratefully acknowledge
a generous grant of telescope time at the SAAO 1.9m.




\clearpage



\clearpage 

\clearpage 

\clearpage 

\clearpage 

\clearpage 

\clearpage 

\clearpage 

\clearpage 

\clearpage 

\clearpage 

\clearpage 

\clearpage 

\clearpage 

\clearpage

\begin{deluxetable}{llrl}
\tabletypesize{\scriptsize}
\tablecaption{Observations\label{obs_tab}}
\tablewidth{0pt}
\tablehead{
\colhead{JD} & \colhead{Date}   & \colhead{N exp}   & \colhead{Comments}
}
\startdata

\multicolumn{4}{c}{1995 SAAO V-band Data}\\
\tableline
2449833	&	25 Apr	&	11	& outburst decline	\\
2449835	&	27 Apr	&	21	& outburst decline	\\
2449837	&	29 Apr	&	19	&	\\
2449838	&	30 Apr	&	18	&	\\
2449839	&	01 May  &	16	&	\\
\tableline
 & & & \\
\multicolumn{4}{c}{2001 Calypso I-band Data}\\
\tableline
2451988	&	20 Mar	&	1	&	\\
2451990	&	22 Mar	&	3	&	\\
2451992	&	24 Mar	&	1	&	\\
2451994	&	26 Mar	&	2	&	\\
2451999	&	31 Mar  &	1	&	\\

2452026	&	27 Apr	&	27	& \\
2452028	&	29 Apr	&	13	& outburst rise? \\

2452050	&	21 May	&	17	&	\\
2452051	&	22 May	&	19	&	\\
2452052	&	23 May	&	23	&	\\
2452053	&	24 May	&	23	&	\\
2452054	&	25 May	&	34	&	\\
2452055	&	26 May	&	29	&	\\
2452056	&	27 May	&	19	&	\\
2452058	&	29 May	&	38	&	\\
2452059	&	30 May	&	52	&	\\
2452060	&	31 May	&	32	&	\\

2452084	&	24 Jun	&	29	&	\\
2452086	&	26 Jun	&	3	&	\\
2452087	&	27 Jun	&	36	& outburst rise	\\
2452088	&	28 Jun	&	23	& outburst rise	\\
2452089	&	29 Jun	&	21	& outburst rise	\\
2452090	&	30 Jun	&	15	& outburst rise	\\
\enddata

\tablecomments{All observations are in quiescence unless otherwise noted.}

\end{deluxetable}

\clearpage

\begin{deluxetable}{rrr}
\tablewidth{0pt}
\tablecaption{Calypso I-band Photometry\label{cal_phot_tab}}
\tablehead{ \colhead{Julian Day}        & \colhead{I}       & \colhead{Error} \\
	    \colhead{(day)}	& \colhead{(mag)}   & \colhead{(mag)} }
\startdata
2451988.938      & 20.34 & 0.03  \\
2451990.986      & 20.05 & 0.04  \\
2451990.995      & 20.11 & 0.05  \\
2451991.002      & 20.17 & 0.06  \\
2451992.882      & 20.13 & 0.03  \\
2451994.914      & 20.18 & 0.03  \\
2451994.925      & 20.24 & 0.03  \\
2451999.937      & 20.34 & 0.06  \\
2452026.752      & 19.67 & 0.04  \\
2452026.760      & 19.93 & 0.04  \\
2452026.767      & 19.88 & 0.03  \\
2452026.774      & 19.96 & 0.03  \\
2452026.781      & 19.92 & 0.03  \\
2452026.789      & 19.91 & 0.03  \\
2452026.804      & 19.90 & 0.03  \\
\enddata

\tablecomments{Table~\ref{cal_phot_tab} is presented in its entirety in the
electronic edition of the Astronomical Journal.  A portion is shown here for 
guidance regarding its form and content. }

\end{deluxetable}

\clearpage

\begin{deluxetable}{rrr}
\tablewidth{0pt}
\tablecaption{SAAO V-band Photometry\label{sa_phot_tab}}
\tablehead{ \colhead{Julian Day}        & \colhead{V}       & \colhead{Error} \\
	    \colhead{(day)}	& \colhead{(mag)}   & \colhead{(mag)} }
\startdata
2449833.436      & 19.78 & 0.04  \\
2449833.447      & 19.81 & 0.04  \\
2449833.459      & 19.83 & 0.04  \\
2449833.469      & 19.81 & 0.03  \\
2449833.500      & 19.82 & 0.04  \\
2449833.530      & 19.85 & 0.04  \\
2449833.560      & 19.83 & 0.03  \\
2449833.573      & 19.83 & 0.04  \\
2449833.594      & 19.84 & 0.04  \\
2449833.615      & 19.87 & 0.05  \\
2449833.634      & 19.90 & 0.05  \\
2449835.374      & 20.53 & 0.04  \\
2449835.386      & 20.57 & 0.04  \\
2449835.397      & 20.67 & 0.04  \\
2449835.408      & 20.68 & 0.05  \\
\enddata

\tablecomments{Table~\ref{sa_phot_tab} is presented in its entirety in the
electronic edition of the Astronomical Journal.  A portion is shown here for 
guidance regarding its form and content. }

\end{deluxetable}






\begin{thebibliography}{}
\bibitem[Baraffe \& Chabrier(1996)]{bar96} Baraffe, I.~\& Chabrier, G.\ 1996, 
	\apjl, 461, L51
\bibitem[Bochkarev, Karitskaya, \& Shakura(1979)]{boc79} Bochkarev, N.~G.,
	Karitskaya, E.~A., \& Shakura, N.~I.\ 1979, SvA, 23, 8
\bibitem[Fabian, Pringle, \& Rees(1975)]{fab75} Fabian, A.~C., Pringle, J.~E., 
	\& Rees, M.~J.\ 1975, \mnras, 172, 15P
\bibitem[Gray(1976)]{gra76} Gray, D. F. 1976, The Observation and Analysis of 
	Stellar Photospheres (New York: Wiley-Interscience) 
\bibitem[Grindlay, Heinke, Edmonds, \& Murray(2001)]{gri01} Grindlay, J.~E., 
	Heinke, C., Edmonds, P.~D., \& Murray, S.~S.\ 2001, Science, 292, 2290
\bibitem[Hakala, Charles, Johnston, \& Verbunt(1997)]{hak97} Hakala, P.~J., 
	Charles, P.~A., Johnston, H.~M., \& Verbunt, F.\ 1997, \mnras, 285, 693
\bibitem[Horne \& Baliunas(1986)]{hor86} Horne, J.~H.~\& Baliunas, S.~L.\ 
	1986, \apj, 302, 757 
\bibitem[Hurley \& Shara(2002)]{hur02} Hurley, J.~R.~\& Shara, M.~M.\ 2002,
	\apj, in press
\bibitem[Hut(1984)]{hut84} Hut, P.\ 1984, \apjs, 55, 301
\bibitem[Knigge et al.(2002)]{kni02} Knigge, C., Shara, M.~M., Zurek, D.~R.,
	Long, K.~S., \& Gilliland, R.~L. 2002, to be published in ASP Conf.
	Ser., Stellar Collisions, Mergers, and their Consequences, 
	astro-ph/0012187
\bibitem[Kukarkin \& Mironov(1970)]{kuk70} Kukarkin, B.~V.~\& Mironov, A.~V.\ 
	1970, \azh, 47, 1211
\bibitem[Margon, Downes, \& Gunn(1981)]{mar81} Margon, B., Downes, R.~A., 
	\& Gunn, J.~E.\ 1981, \apjl, 247, L89
\bibitem[Naylor et al.(1989)]{nay89} Naylor, T.~et al.\ 1989, \mnras, 
	241, 25P
\bibitem[Oosterhoff(1941)]{oos41} Oosterhoff, P.~T.\ 1941, 
	Annalen van de Sterrewacht te Leiden, 17, 41
\bibitem[Pooley et al.(2002)]{poo02} Pooley, D., Lewin, W.~H.~G., Homer, L.,
	Verbunt, F., Anderson, S.~F., Gaensler, B.~M., Margon, B., Miller, J.,
	Fox, D.~W., Kaspi, V.~M., \& van der Klis, M. 2002, \apj, submitted,
	astro-ph/0110192
\bibitem[Press \& Rybicki(1989)]{pre89} Press, W.~H.~\& Rybicki, G.~B.\ 1989, 
	\apj, 338, 277
\bibitem[Sandquist, Bolte, Stetson, \& Hesser(1996)]{san96} Sandquist, E.~L., 
	Bolte, M., Stetson, P.~B., \& Hesser, J.~E.\ 1996, \apj, 470, 910
\bibitem[Scargle(1982)]{sca82} Scargle, J.~D.\ 1982, \apj, 263, 835
\bibitem[Shara, Potter, \& Moffat(1987)]{sha87} Shara, M.~M., Potter, M., 
	\& Moffat, A.~F.~J.\ 1987, \aj, 94, 357
\bibitem[Shara, Potter, \& Moffat(1990)]{sha90} Shara, M.~M., Potter, M., 
	\& Moffat, A.~F.~J.\ 1990, \aj, 99, 1858
\bibitem[Stetson(2000)]{ste00} Stetson, P.~B.\ 2000, \pasp, 112, 925
\bibitem[Tody(1986)]{tod86} Tody, D. 1986, Proc. SPIE, 627, 733
\bibitem[Warner(1995)]{war95} Warner, B.\ 1995,
        Cambridge Astrophysics Series, Cambridge, New York: Cambridge
	        University Press, |c1995
\end{thebibliography}
\end{document}